\documentclass[aps,prb,showpacs,reprint]{revtex4-1}

\usepackage{graphicx}
\usepackage{amsmath}
\usepackage{bbm}

\begin{document}
\title
{Non-ballistic spin separator based on Y-shaped nanostructure with a quantum point contact}
\author{P. W\'ojcik}
\email[Electronic address: ]{pawelwojcik@fis.agh.edu.pl}
\author{M. Wo{\l}oszyn}
\author{B. J. Spisak}
\author{J. Adamowski}
\affiliation{AGH University of Science and Technology, Faculty of
Physics and Applied Computer Science, al. Mickiewicza 30,
Krak\'ow, Poland}

\begin{abstract}
A proposal of a spin separator based on the spin Zeeman effect in Y-shaped nanostructure with
a quantum point contact is presented. Our calculations show that the appropriate tuning of the 
quantum point contact potential and the external magnetic field leads to the spin separation of the
current: electrons with opposite spins flow through the different output branches.
We demonstrate that this effect is robust against the scattering on impurities.
The proposed device can also operate as a spin detector, in which -- depending on the electron spin --
the current flows through one of the output branches.
\end{abstract}


\maketitle

A design of a controllable source of spin-polarized electrons and an efficient injection of
the spin-polarized current into a semiconductor is a basic requirement for a fabrication of novel spintronic
devices. The original idea of the spin transistor\cite{Datta1990} was assumed that spin-polarized 
charges are injected into semiconductor channel from the ferromagnetic contact.  
However, due to the conductivity mismatch between the ferromagnetic and semiconductor materials, the efficiency
of the spin injection at ferromagnet/semiconductor interface is rather low --
experimentally reported at the level of a few percent.\cite{Hammar1999, Schmidt2000, Crooker2005}
The low spin-injection rate results in the low signal in the experimental realization of the
spin transistor.\cite{Koo2009}
In order to overcome the conductivity mismatch, the spin-polarized current is injected into the semiconductor
using the magnetic semiconductors.~\cite{Dietl2000, Jungwirth2006}
In the experiments with the magnetic-semiconductor/semiconductor interface,
the spin polarization rate reaches the value as high as 90~\%.\cite{Fiederling1999,Ohno1999,Jonker2000}
Moreover, the use of magnetic semiconductors in resonant tunneling structures 
allows to construct the spin filter, in which the polarization of the current can be 
changed from fully spin-down to fully spin-up polarized by the bias 
voltage.\cite{Slobodsky2003,Wojcik2012} In our recent paper,\cite{Wojcik2013} we have studied the 
spin filter effect in the resonant tunneling diode  based on ferromagnetic GaMnN
and shown that the spin filter operation can be realized even at room temperature.
Several alternative methods to achieve the spin-polarizing effect in semiconductor nanostructures
have been recently demonstrated in nanowires with spin-orbit interaction,\cite{Nowak2013,Szumniak2012}
quantum dots,\cite{Folk2003} and carbon nanotubes.\cite{Hauptmann2008}
Although all these devices allow to obtain 
the spin polarized current, they do not separate the current into the beams with opposite spin.
This operation is more complicated and requires the application of the device
with at least three terminals: one terminal acts as the input through which the unpolarized current is injected,
and the other two terminals act as outputs, through which the spin polarized beams flow out of the device.
The spin separation effect in the presence of the spin-orbit interaction has been 
theoretically predicted in the ballistic T-shaped\cite{Yamamoto2005} and Y-shaped\cite{Cummings2006,Cummings2008} 
structures as well as quantum rings.\cite{Foldi2006} 
The spin-orbit interaction induced by the quantum point contact (QPC)
has been recently applied to reproduce the Stern-Gerlach experiment in the two dimensional 
electron gas (2DEG).\cite{Ngo2010, Kohda2012} 
\begin{figure}[ht]
\begin{center}
\includegraphics[scale=0.4, angle=0]{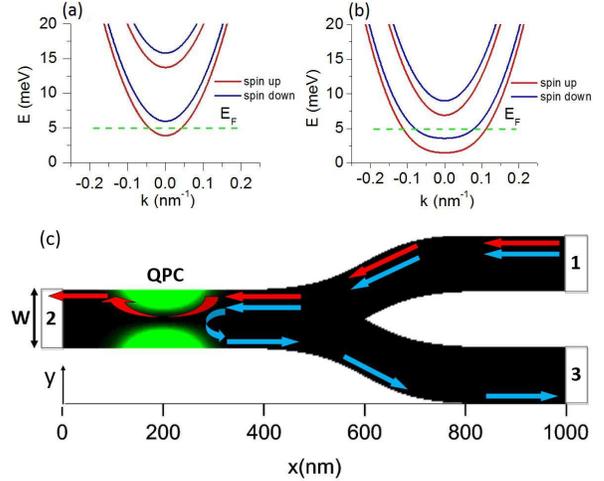}
\caption{(a,b) Dispersion relations $E(k)$ for spin-up (red curves) and spin-down (blue curves) electrons
calculated for (a) the center of the QPC and (b) the leads.
The green horizontal dashed line denotes Fermi energy $E_F$.
(c) Schematic of the Y-shaped nanostructure with the QPC (the green regions illustrate the QPC potential energy
profile [cf. Eq.~(\ref{QPC})]).
The unpolarized current injected from electrode 1 is separated into the two spin-polarized electron beams.
Red (blue) arrows depict the currents of the spin-up (spin-down) electrons.
}
\label{fig1}
\end{center}
\end{figure}
However, the operation of the devices based on the spin-orbit interaction
is very sensitive to the scattering processes and can be affected by the
spin relaxation according to the Dyakonov-Perel mechanism.\cite{Perel1971}
Therefore, the operation of the spin separator exploiting the spin-orbit interaction is restricted to the ballistic regime,
which is a serious obstacle from the experimental point of view.
In this letter, we propose the spin separator based on the Y-shaped 2DEG with the 
QPC located in one of the branches. We show that the proposed nanodevice
can be used both for the separation and detection of the electron spin.
Moreover, if the nanodevice operates in the regime, in which the current is carried through the edge states,
then -- in analogy to the quantum Hall effect -- the spin separation is robust against the scattering.

\begin{figure}[ht]
\begin{center}
\includegraphics[scale=0.45]{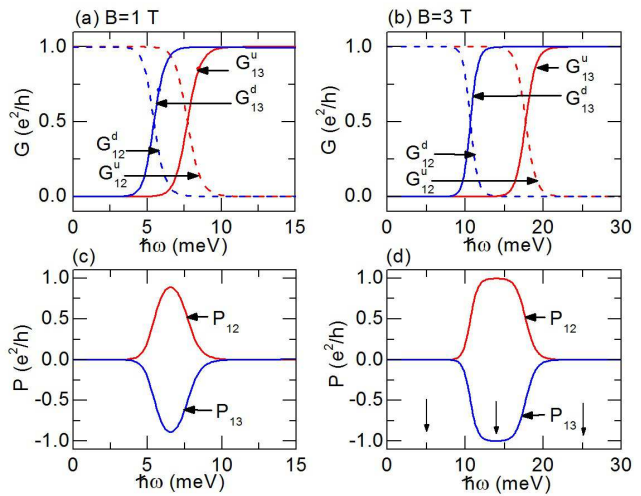}
\caption{Spin-up (red curves) and spin-down (blue curves) conductance $G^{u(d)}_{ij}$ as a function
of confinement energy $\hbar \omega$ for magnetic field (a) $B=1$~T and (b) $B=3$~T.
Panels (c) and (d) show the differences of the conductances
$P_{12}= G^{u}_{12}-G^{d}_{12}$ and $P_{13}=G^{u}_{13}-G^{d}_{13}$.
In panel (d), the vertical arrows mark the values of $\hbar \omega$
chosen to present the results in Fig.~\ref{fig3}.
}
\label{fig2}
\end{center}
\end{figure}

We consider the Y-shaped two-dimensional nanostructure with the QPC located near the contact 2 [Fig.~\ref{fig1}(c)].
Experimentally, similar construction but in the quantum ring geometry 
has been fabricated by the use of the lithography technique.~\cite{Chang2008}
In the presence of the external magnetic field $\mathbf{B}=(0,0,B)$, the Hamiltonian of the system takes on 
the form
\begin{equation}
\hat{H} =\left [ \frac{ (-i\hbar \nabla +e\mathbf{A}) ^2}{2 m_e} + U(\mathbf{r}) \right ] \mathbf{1} 
 + \frac{1}{2}g\mu_B B \sigma _z \;,
\end{equation}
where $\mathbf{A}=(yB,0,0)$ is the vector potential, $m_e$ is the conduction-band mass,
$\mathbf{1}$ is the $2\times 2$ unit matrix, and $\sigma _z$ is the $z$-spin Pauli matrix. 
Potential energy $U(\mathbf{r})=U_c(x,y)+U_{QPC}(x,y)$
is the sum of the Y-shaped confinement potential energy $U_c(x,y)$ (we assume the hard-wall confinement in the $y$ direction)
and the electron potential energy in the QPC
\begin{equation}
U_{QPC}(x,y)=\frac{1}{2} m_e^2 \omega^2 y^2 \exp {\left [ \frac{-(x-x_0)^2}{2 d^2} \right ]} \;,
\label{QPC}
\end{equation}
where $d$ determines the $x$-extension of the QPC, $x_0$ defines its position, and $\hbar \omega$
is the energy of the transverse parabolic confinement in the QPC region.
We assume that the confinement in the $z$ direction is so strong
that electrons occupy the ground-state resulting from the size quantization along this axis.
In the calculations, we adopt the following geometrical parameters: width of the channel $W=40$~nm,
length of the nanostructure $L=1000$~nm, $d=40$~nm, and $x_0=200$~nm.
We use the material parameters corresponding to In$_{0.5}$Ga$_{0.5}$As, 
i.e. $m_e=0.0465m_0$ and $g=-8.97$, however the spin separation effect 
will be observed for any semiconductor material with sufficiently large spin Zeeman splitting.
The numerical calculations have been performed using the tight-binding method on
the square lattice with $\Delta x = \Delta y=2$~nm and the hopping energy $t=\hbar ^2/(2m_e\Delta x ^2)$.
We have used the Kwant package\cite{kwant} to determine the spin-dependent conductance
$G^{u(d)}_{ij}$ between the contacts $i$ and $j$ ($i, j = 1,2,3$, $u$ and $d$ correspond to spin-up and spin-down, respectively).
In the proposed nanostructure electrons are injected from the lead $1$ acting as the input and flow out from 
the device via the leads $2$ and $3$ (see Fig.~\ref{fig1}). 
Figure~\ref{fig2} (upper panels) presents the spin-dependent conductance $G^{u(d)}_{i,j}$ as a function of the QPC
confinement energy $\hbar \omega$ for magnetic field (a) $B=1$~T and (b) $B=3$~T.
We see that the rapid decrease of the conductance between the contacts $1$ and $2$
for the spin-up electrons occurs for the higher energy than for the spin-down electrons.
Simultaneously, for both the spin polarizations the decrease of the electron transmission into the channel $2$
is accompanied by the increase of the transmission into the channel $3$.
This means that in the confinement energy regime, for which the conductance of the spin-up electrons through the channel $2$
is still high, the spin-down electrons are reflected from the QPC and flow through the channel $3$.
The current splits into the two spin-polarized beams.
The splitting effect is quantitatively presented  in Figs.~\ref{fig2}~(c,d)
which show the differences of the conductances $P_{12}= G^{u}_{12}-G^{d}_{12}$ and $P_{13}=G^{u}_{13}-G^{d}_{13}$.
In Figs.~\ref{fig2}~(c,d) the spin separation of the current is revealed as the peak (dip),
which corresponds to the spin-up (spin-down) polarization of the current flowing through the corresponding channel.
\begin{figure}[ht]
\begin{center}
\includegraphics[scale=0.4]{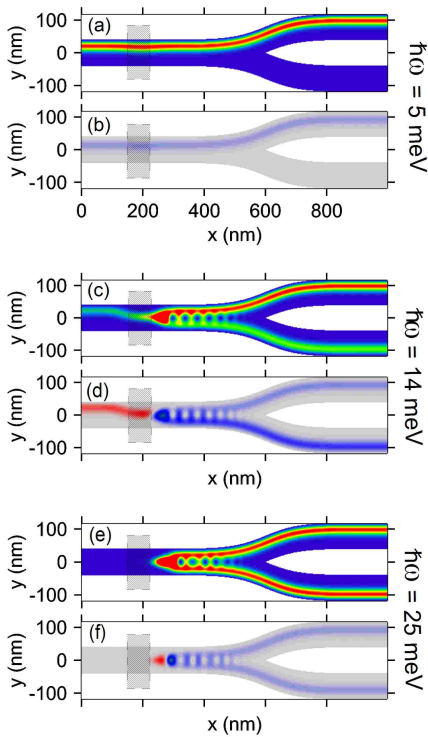}
\caption{Electron density (a,c,e) and spin density (b,d,f) in the nanostructure
for (a,b)  $\hbar \omega=2$~meV, (c,d) $\hbar \omega=14$~meV, and (e,f) $\hbar \omega=25$~meV.
The gray rectangle represents the position of the QPC.}
\label{fig3}
\end{center}
\end{figure}

The spin separation mechanism proposed in this paper results from the joint effect of the spin Zeeman 
splitting and the formation of edge states.
If the magnetic field is applied, the spin degeneration of transverse electron states is lifted by 
the Zeeman effect [cf. the dispersion relations in Fig.~\ref{fig1}~(a,b)].
In the calculations we have adjusted the Fermi level in the leads so that only the two lowest transverse states  
(one corresponding to spin-up and one corresponding to spin-down) are occupied [see Fig.~\ref{fig1}(b)]. 
The spin-up and spin-down electrons with the chosen energy are injected into the system from the lead 1 and flow towards the QPC
located in the channel 2. Due to the spin Zeeman splitting the reflection probabilities in the QPC region for spin-up and spin-down
electrons are different. 
The increase of the confinement energy $\hbar \omega$ leads to the increase of the transverse state energies in the QPC.
In particular, we can tune $\hbar \omega$ so that only the spin-up energy level is located below the Fermi energy.
[see Fig.~\ref{fig1}(a)]. 
The absence of the available spin-down electron states in the QPC region results in a backscattering 
of spin-down electrons.
On the other hand, the spin-up electrons still have a high transmission probability through the QPC.
Therefore, the current splits into the two spin-polarized electron beams.
In order to obtain the full separation of the electrons with opposite spins we exploit
the orbital effect, which causes that the transport of electrons is carried by the edge states
that are formed in a sufficiently high magnetic field.
Due to the orbital effect, the electrons with negative and positive velocities (along the $x$-axis) are  spatially separated.
The electrons flowing in a certain direction are always shifted, by the Lorentz force, to the right boundary of the conductive channel
with respect to the direction of the current flow (cf. Fig.~\ref{fig3}).
This causes that the spin-down electrons backscattered from the QPC are injected into the channel 3. 
In other words, the orbital effect prevents the spin-down polarized electrons reflected from the QPC to flow back into the channel 1.
The resulting spin separation effect is clearly demonstrated in Fig.~\ref{fig3}.  
For $\hbar\omega=5$~meV both the spin-up and spin-down electrons are transmitted through the QPC [Fig.~\ref{fig3}(a)]
and reach the contact 2. We see that the current is partially spin polarized [Fig.~\ref{fig3}(a)], which is a consequence of unequal
electron densities of states at the Fermi level which results from the spin Zeeman effect.
For $\hbar \omega=14$~meV [Fig.~\ref{fig3}(b)] the Y-shaped nanostructure acts as the (almost) perfect spin splitter.
Depending on their spins the electrons injected from the contact 1 are either transmitted through the QPC into the channel 2 or
reflected into the channel 3.
As described above, the backscattering of the spin-down electrons is very strong [cf. Fig.~\ref{fig3}(d)].
We have found that the nanodevice with the parameters of Fig.~\ref{fig3}(d)
is the optimal realization  of the spin separator, in which the electrons with the opposite spins
are spatially separated and leave the nanostructure via the different conduction channels.
The further increase of the confinement energy $\hbar \omega$ [Fig.~\ref{fig3}(e,f)]
causes that in the QPC region  there are no available quantum states for both the spin-up and spin-down electrons.
The electrons with either spin are fully reflected from the QPC and due to the orbital effect flow through the channel 3.

\begin{figure}[ht]
\begin{center}
\includegraphics[scale=0.4]{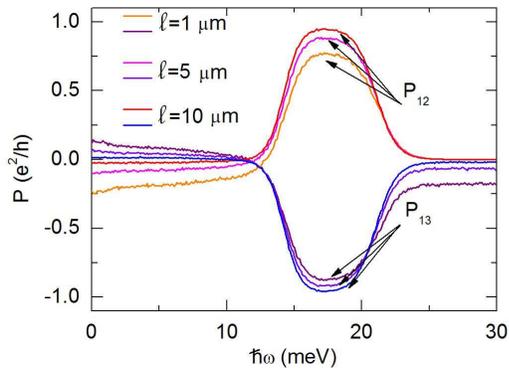}
\caption{Spin polarization $P_{12}$ and $P_{13}$
as a function of the QPC confinement energy $\hbar \omega$.
Results calculated in the non-ballistic regime for different values of mean free path $\ell$.
}
\label{fig4}
\end{center}
\end{figure}

The results presented so far have been obtained with the neglect of the scattering.
Nevertheless, form the experimental point of view it is desirable to construct the spin selector which 
acts in the non-ballistic regime. Now, we will show that the nanodevice proposed in our paper is robust 
against the scattering. Due to the orbital effect, the electrons flowing in a certain channel are moved towards
the edge according to the Lorentz force direction.
For the sufficiently strong magnetic field, the currents flowing in the opposite
directions are transported through the edge states localized at two opposite sides of the channel.
If we increase the magnetic field, the separation between edge states carrying the current in the opposite 
directions increases.
This effect leads to the suppression of the backscattering since the electron can change its momentum only 
if it is scattered from the edge state localized on one side of the channel to that on the other side.
The scattering is suppressed due the vanishingly small overlap between the wave functions localized 
on the opposite sides of the channel.
This mechanism is well known and is the origin of the 'zero' resistance in the quantum Hall effect.\cite{Datta}
We have quantitatively described this effect introducing the model
according to which the spin-independent scattering is included by assuming that the transfer energy $t$
is uniformly distributed within the range $\mathcal{W}/2 < t < \mathcal{W}/2$.\cite{Ando1991}
The relation between the strength of scatterers and the mean free path $\ell$ is given by\cite{Ando1991}
\begin{equation}
 \frac{\mathcal{W}}{E_F}=\left ( \frac{ 6 \lambda _F ^3}{\pi ^3 \Delta x ^2 \ell } \right ) ^{1/2} \;,
\end{equation}
where $\lambda_F$ is the Fermi wavelength.
In our calculations, we have applied the realistic values of the mean free path, which was experimentally
measured to be greater than $1\mu$m for the 2DEG in InGaAs.\cite{Engels1997}
Fig.~\ref{fig4} shows $P_{12}$ and $P_{13}$
calculated in the non-ballistic regime as a function of $\hbar \omega$ for several values of the mean free path.
The results presented in Fig.~\ref{fig4} have been obtained by averaging over $10^4$
computational runs for each value of the energy $\hbar \omega$.
We see that -- in the considered non-ballistic regime -- the scattering
does not affect the spin-splitting effect. The pronounced peak and dip,
which demonstrate spin separation are still clearly visible.
We should emphasize that the insensitivity of the scattering
becomes greater, if the external magnetic field increases. 
Since in this regime the effect of the external magnetic field 
is dominating, we neglect the spin-orbit interaction in our calculations.

In conclusion, we have proposed the non-ballistic spin separator based on the Y-shaped 2DEG with the QPC.
We have shown that by the appropriate tuning of confinement energy in the QPC,
the input unpolarized current can be splitted  into two fully spin-polarized beams,
whereas the electrons with opposite spins flow through the different branches of the nanostructure.
The separation mechanism has been explained  as the joint effect of the spin Zeeman splitting and transport
via the edge states generated in the external magnetic field.
We show that the proposed spin separation mechanism is robust against the scattering.
Although the results have been presented as a function of the QPC confinement 
energy $\hbar \omega$, in the experimental realization this energy can be tuned by changing
the voltage applied to the QPC contacts.
In this structure the spin separation effect can be easily switched on/off by the change of the voltage applied to 
the nearby gate.
It is also worth noting that if the fully spin polarized current is injected from the input electrode,
the current will flow through only one of the output branches.
In this case, when measuring the current in both the output,
we obtain the  information about the spin polarization of the current injected into the system.
This means that the proposed nanodevice can also acts as a detector of the spin polarized current. \\

This work has been supported by the National Science Centre, Poland, under grant DEC-2011/03/B/ST3/00240.

%\bibliography{refs}

\end{document}